\documentclass{PoS}
\usepackage{verbatim}
\usepackage{bm}
\usepackage{graphicx}
\usepackage{amsmath}
\usepackage{amsthm}
\usepackage{amsfonts}
\usepackage{amssymb}
\newcommand{\tht}{\textheight}
\newcommand{\ig}{\includegraphics}

\title{The finite volume spectrum of excited states from lattice QCD simulations}

\ShortTitle{Excited state spectra}

\author{John Bulava\\
 School of Mathematics, 
            Trinity College, Dublin 2, Ireland\\
E-mail: \email{jbulava@tcd.ie}
}

\author{Brendan Fahy, You-Cyuan Jhang, David Lenkner, Colin J. Morningstar\\
        Department of Physics, Carnegie Mellon University, Pittsburgh, PA 15213, U.S.A.\\
        E-mail: \email{colin\_morningstar@cmu.edu, bfahy@andrew.cmu.edu, lenknerd@gmail.com, yjhang@andrew.cmu.edu}}
\author{Justin Foley\\
        Department of Physics, University of Utah, Salt Lake City, UT 84112, U.S.A.\\
        E-mail: \email{justin.foley@gmail.com}}

\author{\speaker{Keisuke J. Juge}\thanks{On leave from Department of Physics, University of the Pacific, Stockton CA, U.S.A.}\\
        Institute of Particle and Nuclear Studies,\\ High Energy Accelerator Research Organization (KEK), Tsukuba, Ibaraki 305-0801, Japan\\
        E-mail: \email{kjuge@pacific.edu}}
\author{Chik Him Wong\\
Department of Physics, University of California, San Diego, CA, U.S.A.\\
E-mail: \email{rickywong@physics.ucsd.edu}}

\abstract{We present results for the spectrum of excited mesons obtained from temporal correlations of spatially-extended single-hadron and multi-hadron operators computed in lattice QCD. The stochastic LapH algorithm is implemented on anisotropic, dynamical lattices for isovectors for pions of mass $390$ MeV. A large correlation matrix with single-particle and two-particle probe operators is diagonalized to identify resonances. The masses of excited states in the $I=1, S=0, T_{1u}^+$ channel as well as the mixing of single and multi-particle probe operators are presented.
}

\FullConference{XV International Conference on Hadron Spectroscopy-Hadron 2013\\
		4-8 November 2013\\
		Nara, Japan }

\begin{document}

\section{Introduction}
The finite volume hadron spectrum determined from lattice QCD simulations provides information concerning hadrons and hadron interactions that supplements the physical properties determined from experiments. In particular, it yields the QCD spectrum (up to cutoff effects) at any quark mass and spatial volume. Physical scattering lengths, phase shifts and decay widths can be deduced from this spectrum \cite{Luscher}. In this contribution, we present progress that has been made towards the goal of determining the finite volume stationary state energies of QCD from lattice QCD simulations (earlier reports can be found in \cite{Bulava:2013dra}) utilizing approximate all-to-all quark propagators to include contributions from disconnected diagrams and explicit two-particle operators. In particular, we present our preliminary spectrum in the $I=1,\ S=0, T_{1u}^+$ symmetry channel and results concerning the identification of the energy levels obtained with $2+1$ dynamical clover-improved, anisotropic lattices \cite{Lin:2008pr} generated by the Hadron Spectrum Collaboration.

\section{Lattice QCD}
The stationary state energies in any particular symmetry channel can be  extracted from Monte-Carlo estimates of Euclidean-space correlation functions $C_{ij}(t)=\langle0|{\mathcal O}_i(t+t_0){\mathcal O}^\dagger_j(t_0)|0\rangle$ estimated at a particular lattice spacing and volume. The probe operators ${\mathcal O}_i$ annihilate/create states with particular quantum numbers. A simple example of a probe operator is $\sum\overline{\psi}({\bf x},t)\gamma_5\psi({\bf x},t)$. We compute the Euclidean-space correlation functions on $2+1$ dynamical, anisotropic lattices with spatial lattice spacings $a_s\simeq0.12$ fm and temporal cutoff $a_t^{-1}\simeq5.6$ GeV. Two different masses at $M_\pi\simeq390$ MeV and $240$ MeV have been generated on two spatial volumes ($24^3$ and $32^3$), although our main results so far are from the $24^3$ lattices ($390$ MeV). 

The spectrum in a finite but large enough volume and small quark masses will have multi-particle thresholds. In order to determine the tower of states in a given symmetry channel, we require interpolating/probe operators that have significant overlap with these two-particle states. This typically requires the use of explicit multi-hadron probe operators, each with some finite momentum, as it is well known that single-particle probe operators couple only weakly to two-particle states. In order to construct a large basis of probe operators which include multi-particle contributions in an efficient manner, we construct our correlation functions in each channel using stochastic LapH quark propagators \cite{stochLapH,stochLapHapp}.

\subsection{Stochastic LapH Quark Propagators}

The stochastic LapH method \cite{stochLapH} is a way of estimating quark propagation from all points on the lattice to all other points on the lattice for smeared quark fields. Probe operators for spectroscopy are typically constructed from smeared quark fields to suppress the high-lying modes present in the fields. An efficient way of evaluating smeared quark propagators has been developed in Ref.~\cite{distillation} whose main idea is to propagate only the low energy modes of the spatial Laplacian operator $\tilde{\Delta}$. 
We shall restrict the eigenmodes of the Laplacian to the lowest lying $112$ modes for the $24^3$ lattice and $264$ modes for the $32^3$ lattice which correspond to a smearing cutoff of $\sigma^2=0.33$. The smeared quark fields are then given by 
$$
\tilde{\psi}_{a\alpha}({\bf x},t)=S_{ab}({\bf x},{\bf y};t)\psi_{b\alpha}({\bf y},t)
$$
where $S$ is the Heaviside-step function $\Theta\left(\sigma^2+\tilde{\Delta}\right)$, $a,b$ are color indices and $\alpha,\beta$ are spin indices. The particular choice of the smearing cutoff parameter is described in Ref.~\cite{stochLapH}.

The above method is then combined with a stochastic approach in order to efficiently evaluate contributions from disconnected diagrams and to simplify the construction of multi-hadron operators/correlation functions. In Ref.~\cite{stochLapH}, stochastic noise was introduced only in the LapH subspace and then diluted \cite{dilution} to reduce the variance in the correlation functions. One of the advantages of using approximate all-to-all propagators, apart from being able to evaluate disconnected diagrams, is that it simplifies the construction of the probe operators allowing a larger basis of operators to be used in the variational analysis.

\subsection{Probe Operators}
We outline the construction of probe operators using stochastic LapH quark propagators below.

\subsubsection{Single-Hadron Operators}
Single-hadron operators are constructed from gauge-covariantly-displaced, LapH-smeared quark fields
$$q^A_{a\alpha j}=D^{(j)}\tilde{\psi}^{(A)}_{a\alpha},\ \ \ \overline{q}^A_{a\alpha j}=\tilde{\overline{\psi}}^{(A)}_{a\alpha}\gamma_4 D^{(j)\dagger}$$
as building blocks. Here, $a$ is the color index, $\alpha$ is the Dirac spin index, $A$ is the quark flavor (or quark line), $\gamma_4$ is the temporal Dirac $\gamma$-matrix and the displacement is a product of stout-smeared link variables 
$$
D^{(j)}({\bf x},{\bf x}^\prime)=\tilde{U}_{j_1}({\bf x})\tilde{U}_{j_2}({\bf x}+{\bf d}_2)\cdots\tilde{U}_{j_p}({\bf x}+{\bf d}_{p+1}).
$$
\noindent For single-meson operators, we first form gauge-invariant two-quark elemental operators of the form
$$\overline{\Phi}^{AB}_{\alpha\beta}({\bf p},t)=\sum_{\bf x}e^{i{\bf p}\cdot({\bf x}+\frac{1}{2}({\bf d}_\alpha+{\bf d}_\beta))}\delta_{ab}\overline{q}^{B}_{b\beta}({\bf x},t)q^A_{a\alpha}({\bf x},t)
$$
where ${\bf p}$ is the three-momentum of the meson and $\alpha,\beta$ are the combined spin and displacement indices. The displacements of the fields are needed to ensure the proper G-parity transformations for non-zero displacements ${\bf d}_\alpha$ and ${\bf d}_\beta$ of the LapH smeared quark fields $\tilde{\psi}$. Single baryon elemental operators are constructed in a similar way using three independent smeared and displaced noise vectors $\overline{q}^A_{a\alpha}$. 

The elemental operators must be combined to transform irreducibly under the symmetries of the three-dimensional cubic lattice \cite{Basak:2005ir}. The group theory coefficients $c^{(i)}_{\alpha\beta}$ are used to project out particular irreps, where $i$ is a compound index comprised of three-momentum ${\bf p}$, the irrep $\Lambda$ of the little group of ${\bf p}$, the row $\lambda$ of the irrep, total isospin $I$, third component of isospin $I_3$, strangeness $S$ and the identifier labeling the particular operator in each symmetry channel. The single-hadron source operators then take the form,
$$
\bar{\mathcal M}_i({\bf p},t)=c^{(i)*}_{\alpha\beta}\overline{\Phi}^{AB}_{\alpha\beta}({\bf p},t),\ \ \bar{\mathcal B}_i(t)=c^{(i)*}_{\alpha\beta\gamma}\overline{\Phi}^{ABC}_{\alpha\beta\gamma}({\bf p},t).
$$ 
\subsubsection{Two-Hadron Operators}
Two-hadron operators are formed by combining single-hadron operators with finite momenta with the appropriate group theory coefficients. We require the coefficients for two-hadron operators with total momentum ${\bf p}={\bf p}_a+{\bf p}_b$, isospin $(I,\ I_3)$, strangeness $S$, little group irrep $\Lambda$ and row $\lambda$. The two-hadron operator also carries another index which specifies the particular type of the two-particle operator.

We could construct single-hadron operators with any momentum from the LapH quark propagators, but we have focused on the simplest momenta types (on-axis, planar-diagonal, cubic-diagonal) for most of the symmetry channels for practical reasons. The general form of two-particle operators are given by,
$$
c^{I_{3a}I_{3b}}_{{\bf p}_a\lambda_a;{\bf p}_b\lambda_b}{\mathcal H}^{I_aI_{3a}S_a}_{{\bf p}_a\Lambda_a\lambda_ai_a}{\mathcal H}^{I_bI_{3b}S_b}_{{\bf p}_b\Lambda_b\lambda_b}
$$
where ${\cal H}^{I_aI_{3a}S_a}_{{\bf p}_a\Lambda_a\lambda_ai_a}$ is a meson or baryon operator. 
The group theoretical projections of these operators involves the rotations of finite momentum operators requiring us to define the finite momentum states in an arbitrary but a fixed way. We choose a reference direction ${\bf p}_{\rm ref}$ for each class of momentum directions, and for each ${\bf p}$, we select one reference rotation $R^{\bf p}_{\rm ref}$ that transforms ${\bf p}_{\rm ref}$ into ${\bf p}$ as described in Ref.~\cite{stochLapH}. 

\section{Correlation Functions}
Correlation functions, ${\cal C}_{ij}(t)=\langle0|{\cal O}_i(t+t_0)\overline{\cal O}_j(t_0)|0\rangle$ involving the various gauge invariant, single/two-hadron operators are expressed in terms of path integrals which are estimated using the Monte-Carlo method. We normalize the operators at some early timeslice $\tau_N$ and form 
$\hat{C}_{ij}(t)={\cal C}_{ij}(t)\left[{\cal C}_{ii}(\tau_N){\cal C}_{jj}(\tau_N)\right]^{-1/2}.$ 
This removes the effects of different normalizations of the operators. 
We then evaluate the eigenvalues of $\hat{C}$ and remove those eigenvectors which have negative eigenvalues and eigenvalues less than a threshold value, which is chosen to be the ratio of the largest eigenvalue over some chosen maximum condition number. The resulting correlation matrix $C(t)$ is then used in the variational analysis. 

\subsection{Analysis}
We use the fixed-coefficient method to rotate our correlation matrix into a diagonal matrix at a specified timeslice, $\tau_D$ for a fixed metric timeslice, $\tau_0$. The metric timeslice and the diagonalization point are chosen to be as large as possible while keeping the statistical noise manageable to perform stable exponential fits. The resultant correlators are given by,
\begin{equation}
 G(t) = U^\dagger\ C(\tau_0)^{-1/2}\ C(t)\ C(\tau_0)^{-1/2}\ U,
\label{eq:rotatedcorr}
\end{equation}
where the columns of $U$ are the orthonormalized eigenvectors of $C(\tau_0)^{-1/2}\ C(\tau_D)\ C(\tau_0)^{-1/2}$. 
We then perform single exponential fits $A_n(e^{-E_n\,t}+e^{-E_n\,(T-t)})$ where $T$ is the temporal extent of the lattice, yielding the energies $E_n$ and the overlaps $A_n$ to the rotated operators for each $n$. The overlaps in the original basis is given by rotating back, $Z^{(n)}_j=C(\tau_0)^{1/2}_{jk}\ U_{kn}\ A_n$ (no summation over $n$) where the indices correspond to the rows and columns of the correlation matrix $C(t)$. 
\begin{figure}[t]
	\begin{center}
  \begin{tabular}{ccc}
    \hspace*{-1cm}
		\begin{minipage}{50mm}
			\begin{center}
      \ig[width=0.25\tht,angle=0]{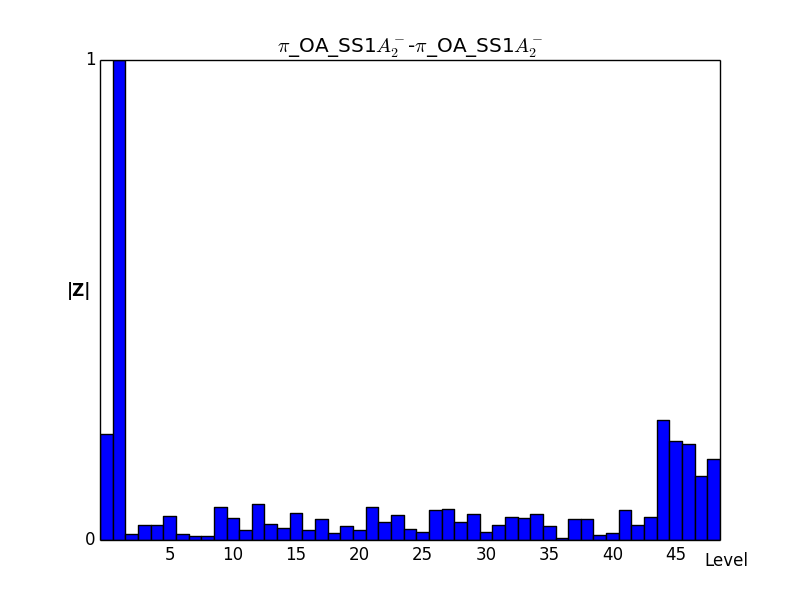}
			\end{center}
		\end{minipage}&
    \hspace*{-3mm}
		\begin{minipage}{50mm}
			\begin{center}
      \ig[width=0.25\tht,angle=0]{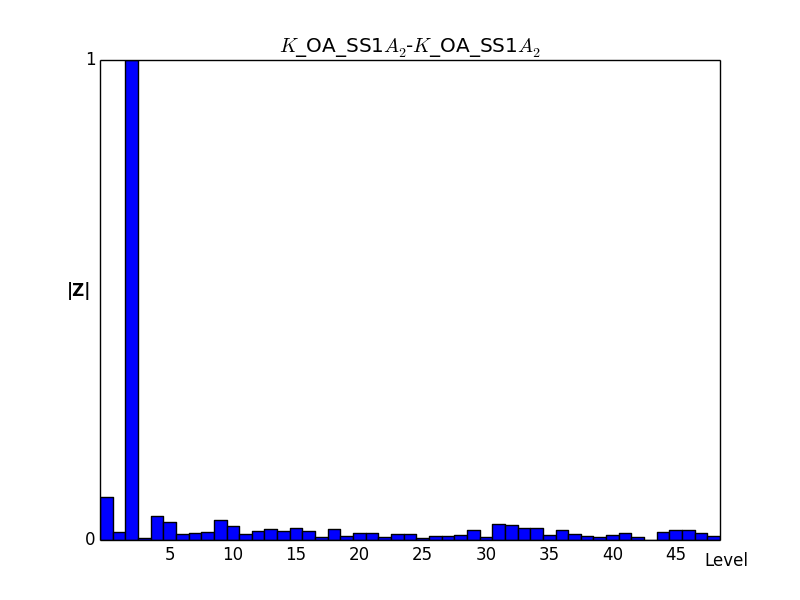}
			\end{center}
		\end{minipage}&
    \hspace*{-3mm}
		\begin{minipage}{50mm}
			\begin{center}	
      \ig[width=0.25\tht,angle=0]{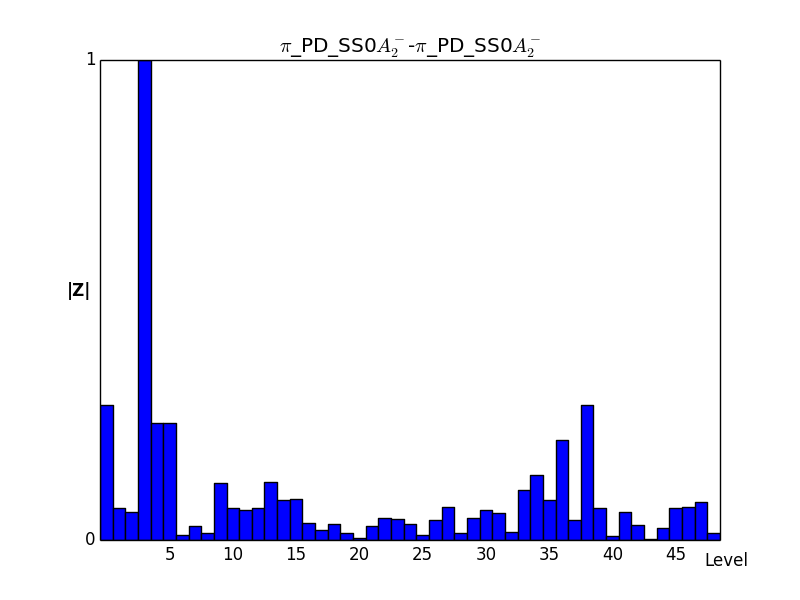}
			\end{center}
		\end{minipage}    
	\end{tabular}
  \label{fig:ov}
  \caption{The overlap factors for three different, two-particle operators plotted against the energy levels $n$. The left and right figures are for $\pi\textrm{-}\pi$-type operators and the middle is for a $K\textrm{-}\bar{K}$-type operator. One finds that the $n=1$ and $n=3$ excited state is predominatly a $\pi{\textrm -}\pi$ state and the $n=2$ state is a $K{\textrm -}\bar{K}$ state.}
	\end{center}
\end{figure}
\subsection{State Identifications}
Identifying the extracted energy levels with a particular physical state is non-trivial, especially when the finite volume spectrum is limited to a couple of different box sizes. The kinematics of our lattices are such that the ground state is a single-particle state, which we identify with state that becomes the $\rho$ resonance in the physical spectrum. Inspection of the overlap factors provides support for the ground state as a single-particle state (not shown).

The largest relative overlaps with the first-excited level come from the explicit $\pi\textrm{-}\pi$ operators with finite momentum in our scenario (Fig.~1). The pion mass and the box sizes used here are such that this state is heavier than the ground state $\rho$ resonance. We expect that the overlap of the two-pion operator to become larger than the single-particle operator for the ground state once the kinematics allows smaller momenta for lighter pions. 

The largest relative overlaps with the second-excited level come from the explicit $K\textrm{-}\bar{K}$ state in this volume and $m_\pi$ (middle plot in Fig.~1). The third level (right figure) is again identified with a $\pi{\textrm -}\pi$ state. Some of the higher levels receive contributions from many operators, but one can proceed in this fashion to try and give the dominating physical content of each level. 

In order to identify the levels extracted with the physical resonances  determined in experiments, we re-diagonalize the correlation matrix using only the single-hadron probe operators. The procedure is similar to the one used for the full matrix. We have determined five of the lowest lying single-hadron states in this way in the $I=1,\ S=0,\ T_{1u}^+$ channel. The energies are plotted in Fig.~2 where the masses are expressed in units of $3/5$ of the $\Omega$ baryon mass. Comparison with the experimental spectrum (right column) shows that the qualitative features of the spectrum agree and that all of the states below $\rho(1700)$ have been determined with the correct level ordering. 

\begin{figure}[t]
\begin{center}
\label{fig:rhospec}
\ig[height=0.28\tht,angle=0]{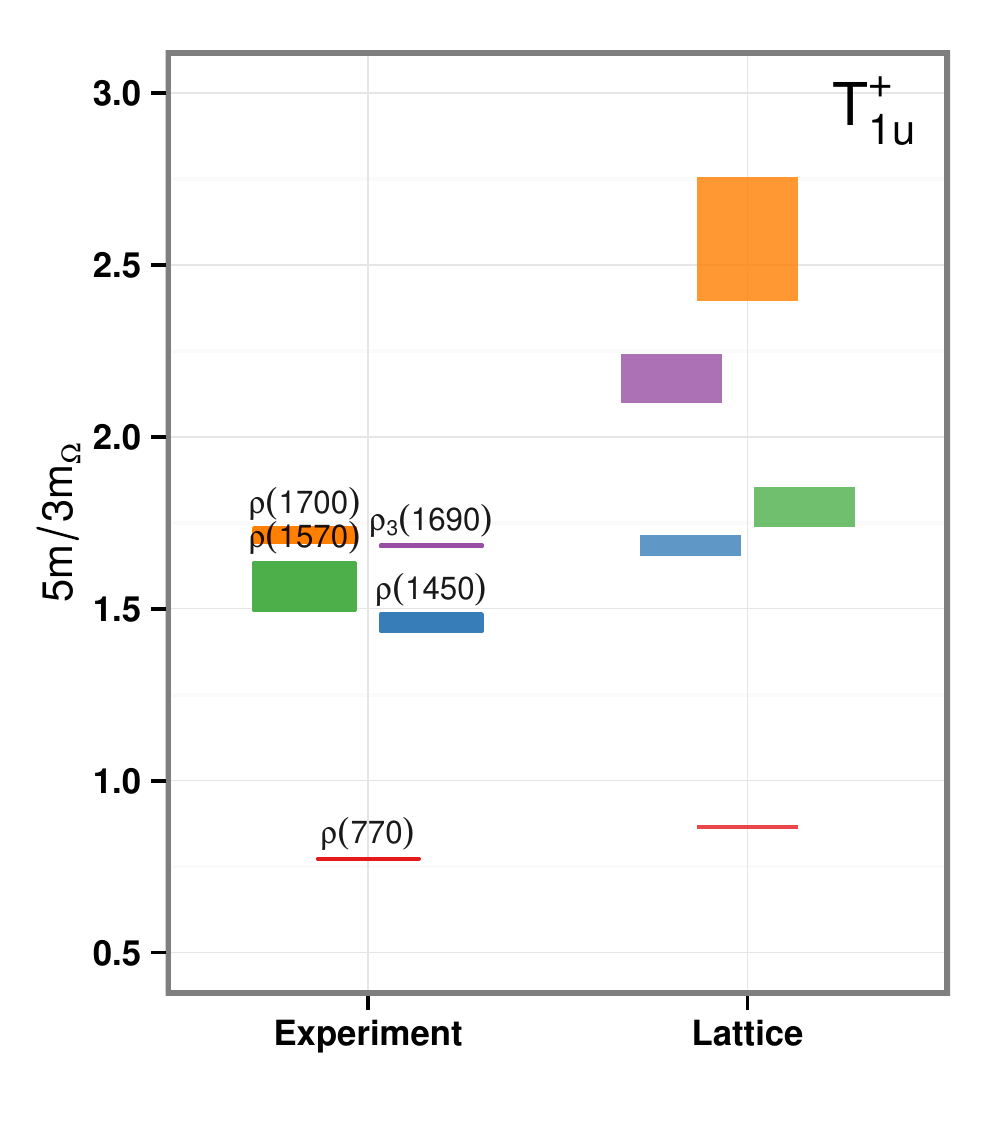}
\caption{Preliminary results for the $I=1,\ S=0$ ($T^+_{1u}$) channel stationary states expected to evolve into the single-meson resonances in infinite volume computed using a $12\times12$ correlation matrix for the ($24^3$|$390$) ensemble.}
\end{center}
\end{figure}

\section{Conclusions}
Progress in our efforts to compute the finite-volume low-lying spectrum of QCD was presented with the focus on the $\rho$ channel. Our approach relies on group theoretical construction of probe operators whose elements are the sources and solutions of stochastic LapH quark propagators. The advantages of the method are that disconnected diagrams and same-time diagrams can be estimated as well as the fact that multi-hadron operators can be formed easily for inclusion in a large variational basis of operators. Preliminary results in the isospin-1, strangeness-0 sector have been presented for the $24^3$ lattice. Analysis of the $32^3$ results for the scattering lengths, phase shift and decay width calculations as well as the lighter pion mass data are in progress and will be presented in future work.

\section*{Acknowledgements}

This work was supported by the U.S. NSF under awards PHY-0510020, PHY-0653315, PHY-0704171, PHY-0969863, PHY-0970137, PHY-1318220 and through TeraGrid/XSEDE resources provided by TACC and NICS under grant numbers TG-PHY100027 and TG-MCA075017.

\end{document}